\documentclass[reprint,superscriptaddress,amsmath,amssymb,aps,prl,twocolumn, nobibnotes]{revtex4-2}

\usepackage{graphicx}
\usepackage{dcolumn}
\usepackage{bm}
\usepackage{color}
\usepackage{amsmath}
\usepackage{breqn}
\usepackage{float}
\usepackage{setspace}
\usepackage{ragged2e}
\usepackage{titlesec}
\usepackage[dvipsnames]{xcolor}
\usepackage[labelfont={color=Black}]{caption}

\begin{document}

\title[]{Interfacial Heat Transport via Evanescent Radiation by Hot Electrons}

\author{William D. Hutchins}
\thanks{These two authors contributed equally.}
\author{Saman Zare}
\thanks{These two authors contributed equally.}
\affiliation{$Department~of~Mechanical~and~Aerospace~Engineering,~University~of~Virginia,~Charlottesville,~Virginia~22904, USA$}

\author{Mehran Habibzadeh}
\author{Sheila Edalatpour}
\affiliation{$Department~of~Mechanical~Engineering,~University~of~Maine,~Orono,~Maine~04469, USA$}

\author{Patrick E. Hopkins}
\email{Corresponding Author: phopkins@virginia.edu}
\affiliation{$Department~of~Mechanical~and~Aerospace~Engineering,~University~of~Virginia,~Charlottesville,~Virginia~22904, USA$}
\affiliation{$Department~of~Materials~Science~and~Engineering,~University~of~Virginia,~Charlottesville,~Virginia~22904, USA$}
\affiliation{$Department~of~Physics,~University~of~Virginia,~Charlottesville,~Virginia~22904, USA$}

\begin{abstract}

We predict an additional thermal transport pathway across metal/non-metal interfaces with large electron-phonon non-equilibrium via evanescent radiative heat transfer. In such systems, electron scattering processes vary drastically and  can be leveraged to guide heat across interfaces via radiative heat transport without engaging the lattice directly. We employ the formalism of fluctuational electrodynamics to simulate the spectral radiative heat flux across the interface of a metal film and a non-metal substrate. We find that the radiative conductance can exceed 300 MW m$^{-2}$ K$^{-1}$  at an electron temperature of 5000 K for an emitting tungsten film on a hexagonal boron nitride substrate, becoming comparable to its conductive counterpart. This allows for a more holistic approach to the heat flow across interfaces, accounting for electron-phonon non-equilibrium and ultrafast near-field phonon-polariton coupling.

\end{abstract}

\maketitle
\onecolumngrid

\twocolumngrid

The thermal boundary conductance (TBC) between two solids relates the heat flux, $q$, to the temperature drop $\Delta T$ across the interface. 
Over the past few decades, theories that describe the interactions among electrons and phonons at interfaces have elucidated various fundamental carrier scattering and conversion processes that drive these interfacial thermal transport pathways~\cite{swartzThermalBoundaryResistance1989a,hopkinsThermalTransportSolid2013a,giriElectronPhononCoupling2020,monachonThermalBoundaryConductance2016,chenInterfacialThermalResistance2022,giriUltrafastNanoscaleEnergy2023}.

One of the most fundamental and ubiquitous transfer mechanisms, thermal radiation, has been neglected in this interfacial heat transfer discussion and concomitant thermal boundary conductance theories. The relatively small fluxes inhibited by the blackbody limit seem negligible when compared to traditionally studied conductive pathways driven by electrons and phonons. In recent years, there has been a renewed interest in this field due to the verification of the prediction of  super Planckian enhancement due to the contribution of evanescent modes~\cite{biehsHyperbolicMetamaterialsAnalog2012a,guoFluctuationalElectrodynamicsHyperbolic2014, principiSuperPlanckianElectronCooling2017}. This is of special importance in the `near-field' regime, when the separation distances are smaller than the thermal wavelength~\cite{cravalhoEffectSmallSpacings1967}. The contribution of evanescent radiative modes can be the dominant thermal transport mechanism within these distances~\cite{chiloyanTransitionNearfieldThermal2015a}. With the ability of evanescent radiative heat transfer (RHT) to exceed the blackbody limit, many experiments have been performed for a range of different geometries, materials, and  gaps ranging from micrometers down to nanometers~\cite{zhangExperimentsNearfieldRadiative2023, ghashamiExperimentalExplorationNearfield2020}. Due to this new capability, there is  growing interest in the effect that evanescent RHT can have on various thermal technologies such as thermophotovoltaics (TPVs)~\cite{lenertNanophotonicSolarThermophotovoltaic2014}, heat-assisted magnetic recording (HAMR) devices~\cite{challenerHeatassistedMagneticRecording2009,stipeMagneticRecording152010}, scanning thermal microscopy~\cite{dewildeThermalRadiationScanning2006, kittelNearfieldThermalImaging2008}, and coherent thermal sources~\cite{greffetCoherentEmissionLight2002b, jonesThermalNearfieldCoherence2013, carminatiNearFieldEffectsSpatial1999}. Hence, we must envelop our standard thermal theory of interfacial transport~\cite{swartzThermalBoundaryResistance1989a} to leverage evanescent RHT in the solid state.

The largest RHT enhancements in natural materials have been reported in polar dielectrics, such as SiC, SiO${_2}$, and hBN, where phonon polariton (PhPs) dominate the evanescent modes~\cite{muletEnhancedRadiativeHeat2002, iizukaAnalyticalTreatmentNearfield2015} in the near field as predicted via Rytov's formalism of fluctuational electrodynamics~\cite{francoeurSpectralTuningNearfield2010, rytovPrinciplesStatisticalRadiophysics1987}. This theory predicts thermal radiation mediated by all propagative, frustrated and surface modes at an arbitrary distance from an emitting body. The frustrated and surface modes are not allowed to propagate outside the emitting body (i.e., they are evanescent), as the parallel component of the wavevector ($k_\rho$) for these modes is greater than the free-space wavevector ($k_0$). However, when another medium is brought close to this body, the evanescent modes can be coupled to the receiving system, and the heat is transported radiatively. The interfaces adjacent to polar dielectric systems are also the source of major thermal boundary resistances due to the large phononic mismatches that occur at metal/dielectric interfaces in devices~\cite{giriReviewExperimentalComputational2020,giriUltrafastNanoscaleEnergy2023}, thus heavily reducing the possible thermal management at these scales. To compound this, the electrons of metal interconnects are often an order of magnitude hotter than the phonon subsystem during operation~\cite{minamitaniInitioAnalysisInitial2021}. Capitalizing on evanescent RHT, the thermal boundary conductance across metal/dielectric interfaces can be tuned by using the electron-phonon non-equilibrium. 

In this Letter, we model the transduction of thermal radiation from a metallic emitter into an insulating dielectric substrate, specifically under the case of strong electron-phonon non-equilibrium. By leveraging the formalism of fluctuational electrodynamics (FED) as well as the classic Drude-Sommerfeld theory of free electrons \cite{ashcroftSolidStatePhysics1976}, we also examine the effect of electron temperature in the metallic thin film on interfacial radiative conductance, $h_{rad}$, in the presence of polaritonic and hyperbolic insulators. By doing so, we demonstrate that there can 

\newpage

\onecolumngrid

\begin{figure}
 \includegraphics[width=\textwidth]{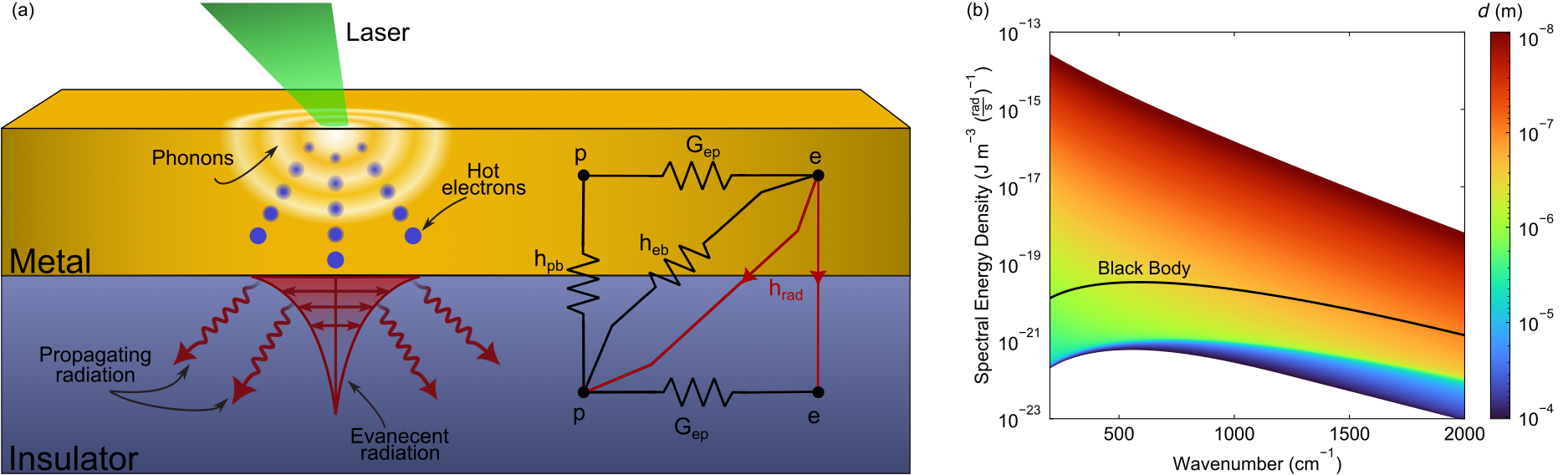}
    \captionsetup{justification=RaggedRight,singlelinecheck=false}    
    \caption[] {(a) Diagram of interfacial transport phenomena. The schematic depicts energy transfer processes between electrons ($e$), phonons ($p$), and boundary ($b$) immediately after an ultrashort laser pulse is absorbed. The evanescent and propagating radiation due to ballistic hot electrons is absorbed into the substrate based on the dielectric properties beneath the interface. (b) The spectral energy density due to radiation of propagating and evanescent modes from a gold film at a distance $d$ (shown in the color bar) away from the surface calculated using FED. The solid black line shows the far-field blackbody limit.}
  \label{fig:1}
\end{figure}

\twocolumngrid

\noindent be more than two-fold increases in overall thermal interfacial transport at high electron temperatures due to the contribution of evanescent radiative heat flux at metal-insulator interfaces. 

The configuration studied here is schematically depicted in Fig. \ref{fig:1}a, where non-equilibrium is illustrated as the separation in thermal energy immediately after excitation from short pulsed laser absorption. Another example of this non-equilibrium occurs in the gate of high-frequency electronic devices~\cite{caiNonequilibriumElectronphononScattering1986b,kralElectrophononResonanceGa1994, carminatiNearFieldEffectsSpatial1999}. These events can cause the electronic bath at temperatures of thousands of Kelvin while maintaining a cold lattice. This non-equilibrium manifests as the opening up of the Fermi surfaces, resulting in an increase in the population and the scattering rates of conducting electrons~\cite{linElectronphononCouplingElectron2008a,giriElectronPhononCoupling2020, giriExperimentalEvidenceExcited2015, hopkinsContributionDbandElectrons2009}.  These increases open up possible radiative photonic states in both the propagating and evanescent regions of the photonic dispersion. The proposed mechanism of RHT across interfaces is fueled by the absorption in insulating dielectrics.  These dielectric materials with high indices can shuttle the evanescent modes from the metal into high-wavevector photonic states, enhancing thermal transport. 

All the processes of the thermal transfer across the interface are depicted in Fig. \ref{fig:1}a, where `e' and `p' refer to the electronic and phononic subsystems, respectively. Each material has its electron-phonon coupling rate, $G_{ep}$ \cite{anisimovElectronEmissionMetal1974}, at which electronic energy is converted to phonon energy. The conductance across the interface includes the phonon contribution to the boundary resistance ($h_{pb}$)~\cite{swartzThermalBoundaryResistance1989a} and the contribution associated with electron collisions at the interface producing phonons in the insulator ($h_{eb}$) \cite{giriInfluenceHotElectron2014}. Finally, there exists a conductance associated with the mechanism proposed by this work, where the thermal radiation emitted by the electrons may be absorbed into the phonons or electrons in the substrate ($h_{rad}$). To represent the scale of available energy contained in thermally excited evanescent modes of a metallic emitter, we calculate the energy density at a distance $d$ away from a gold layer with bulk properties (Fig. \ref{fig:1}b) using FED \cite{joulainSurfaceElectromagneticWaves2005, rytovPrinciplesStatisticalRadiophysics1987, guoFluctuationalElectrodynamicsHyperbolic2014}. As $d$ decreases, the emitted flux increases due to the evanescent contribution such that it exceeds the blackbody limit by several orders of magnitude.  

The formalism for thermal emission by a system with non-equilibrium carriers was provided by Greffet \textit{et al.} \cite{greffetLightEmissionNonequilibrium2018a} asserting that the subsystem temperature of a carrier in non-equilibrium should be used along with its contribution to the dielectric function. Thus, the emission of the metallic layer is dominated by the free-electron contribution during non-equilibrium and before electron-phonon thermalization (i.e., at timescales less than $\sim$ 10 ps after an electronic excitation). At this short timescale, there is an opportunity to sink this heat before the bulk system reaches a high-entropy equilibrium dominated by diffusive processes. The dielectric behavior of free electrons is described by the Drude model as
\begin{equation}
    {\varepsilon}_m(\omega) = 1 - \frac{\omega_p^{2}}{\omega^2 + i\omega \Gamma_{tot}} 
\end{equation}

\noindent where $\omega_p$ is the plasma frequency and $\Gamma_{tot}$ is the total electronic scattering term. As the Drude model predicts a broadband dielectric function across the infrared region of the spectrum (where most of the RHT occurs), the spectral energy density due to thermal emission by the metallic layer also exhibits a broadband behavior, as shown in Fig. \ref{fig:1}b for gold as an example.

\newpage 

\begin{figure}[h!]

 \includegraphics[width=0.5\textwidth]{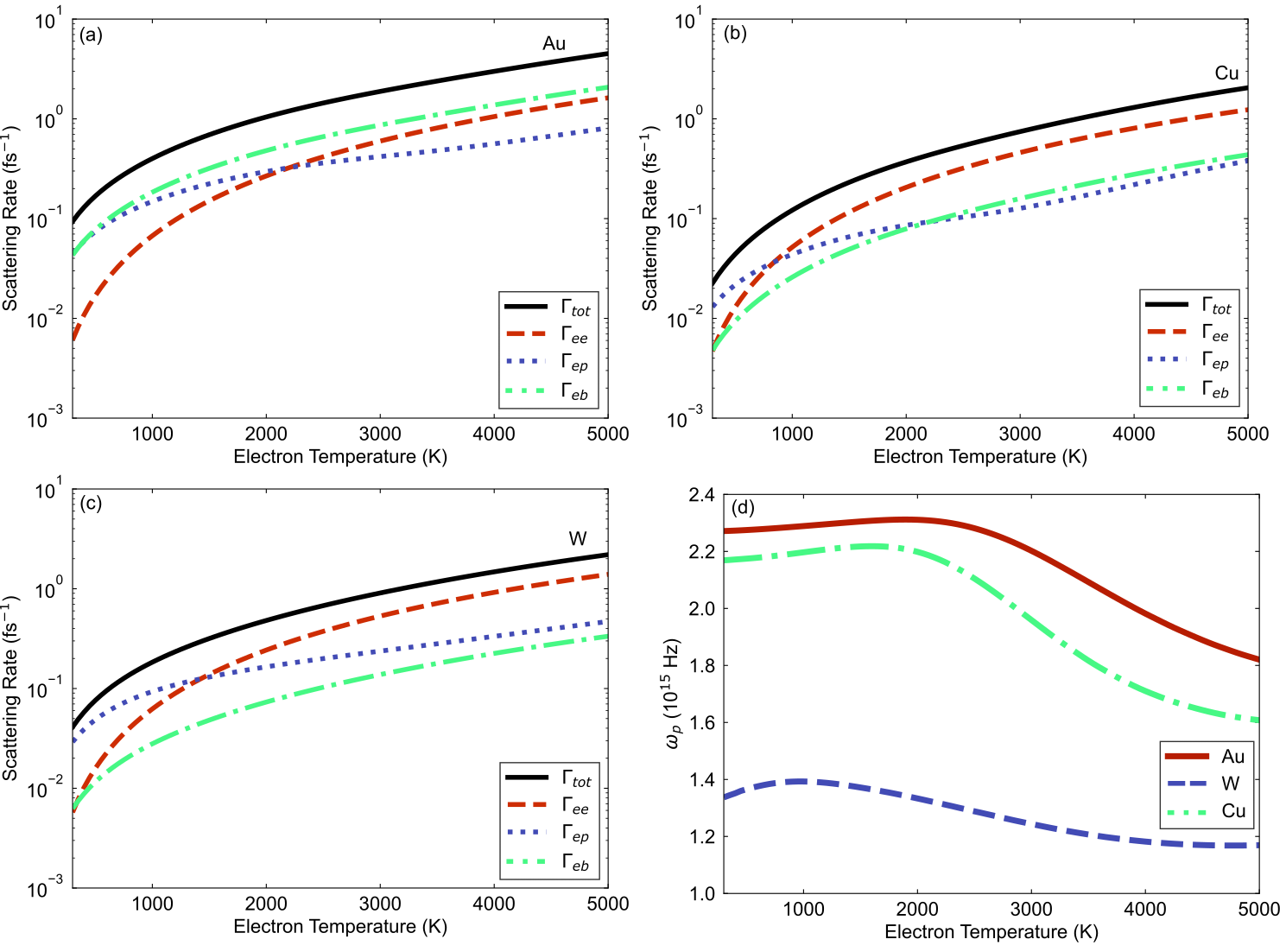}
    \captionsetup{justification=RaggedRight,singlelinecheck=false}\caption[]{(a)-(c) Electron-electron, electron-phonon, electron-boundary, and total scattering rates in a film of gold, copper, and tungsten, respectively. The thickness of the film is set to 10 nm. (d)  Electron temperature trend of the plasma frequency for each metal film. }
  \label{fig:2}
\end{figure}

When considering a non-equilibrium system with elevated electron temperatures, we need to investigate the scope of the Drude oscillator parameters (i.e., $\omega_p$ and $\Gamma_{tot}$) which varies noticeably with electron temperature. The strength of the Drude oscillator is given by the plasma frequency defined as
\begin{equation}
     \omega_p(T_e) = \sqrt{\frac{n_e(T_e) e^2}{ 4 \pi m^*(T_e) \epsilon_0}}
\end{equation}
where $e$ and $\epsilon_0$ are the elementary charge and free-space permittivity, respectively. The parameters which depend on the electron temperature, $T_e$, are the effective mass ($m^*$) and the number density of electrons ($n_e$) given by integrating the occupied density of electronic states over all energies.  The temperature dependencies of $m^*$ and $n_e$ can be computed from the total electron density of states (eDOS) and the chemical potential, $\mu$, detailed in the Supplemental Materials~\cite{supp}\nocite{linElectronphononCouplingElectron2008a, kresseTheoryCrystalStructures1994, anisimovElectronEmissionMetal1974, haynesCRCHandbookChemistry2016,palikHandbookOpticalConstants1998, caldwellSubdiffractionalVolumeconfinedPolaritons2014}. 

There is also a sizable increase in the total electronic scattering rate, $\Gamma_{tot}$, at high temperatures due to the increase in average collisions experienced by electrons at the broadened Fermi surface. The classic Drude model assumes independent electrons, but at the nanoscale, elastic collisions between electrons and boundaries become significant. To account for this effect, we extend the Drude model using Matthiessen's rule~\cite{islamEvaluatingSizeEffects2024}, providing a more accurate total relaxation time, $\Gamma_{\mathrm{tot}}$, as
\begin{equation}
  \Gamma_{tot} =\Gamma_{ee} + \Gamma_{ep} +\Gamma_{eb}
  \end{equation}
where $\Gamma_{ee}$, $\Gamma_{ep}$, and $\Gamma_{eb}$ are the electron-electron, electron-phonon, and electron-boundary scattering rates, respectively. Each of these rates can be calculated from first principles under the theories of Fermi liquid theory (FLT)~\cite{gasparovElectronphononElectronelectronElectronsurface1993a, pinesCollectiveDescriptionElectron1952}, electronic Cerenkov radiation of sound waves~\cite{kaganovRelaxationElectronsCrystalline1957}, and inelastic electron scattering from a vibrating boundary~\cite{sergeevElectronicKapitzaConductance1998}, with the inputs taken from refs.~\cite{linElectronphononCouplingElectron2008a, giannozziQUANTUMESPRESSOModular2009} (see Supplemental Materials for details~\cite{supp}\nocite{linElectronphononCouplingElectron2008a, kresseTheoryCrystalStructures1994, anisimovElectronEmissionMetal1974, haynesCRCHandbookChemistry2016,palikHandbookOpticalConstants1998, caldwellSubdiffractionalVolumeconfinedPolaritons2014}). Throughout the simulations performed in this work, we consider a metal film thickness of 10 nm with a `cold' lattice, i.e., $T_p = \mathrm{300}$ K. This assumption represents the initial transient phase of strong electron-phonon non-equilibrium following an intense electronic excitation. During this scenario, $T_e$ can reach thousands of Kelvin while $T_p$ remains nearly constant on the timescale of picoseconds due to the relatively slow electron-phonon energy exchange. This isolates the impact of hot electrons on radiative heat flux in non-equilibrium conditions, such as after ultrafast laser excitation~\cite{hohlfeldElectronLatticeDynamics2000a} or during high-power electronic device operation~\cite{yngvessonMicrowaveSemiconductorDevices1991,minamitaniInitioAnalysisInitial2021}.  

The temperature dependence of these scattering rates along with the total scattering rate in three representative metals (i.e., Au, Cu, and W) are shown in Figure \ref{fig:2}a-c. At the low-temperature limit of $T_e\sim \mathrm{300}$ K, the dominant mechanism is electron-phonon scattering as the system is close to equilibrium and the phonons scatter more readily since they have higher heat capacity than the free electrons. At electron temperatures higher than $\mathrm{2500}$ K, this trend inverts and the self-interaction of the electronic sea, i.e., the electron-electron scattering, dominates over the electron-phonon counterpart. While all three metal films show similar electron-electron scattering rates, the gold film exhibits a higher electron-phonon scattering rate than copper and tungsten. Although gold has the lowest electron-phonon coupling among these three metals~\cite{linElectronphononCouplingElectron2008a}, the electron-phonon scattering rate is the highest since gold has the lowest sound speed (see Supplemental Materials~\cite{supp}\nocite{linElectronphononCouplingElectron2008a, kresseTheoryCrystalStructures1994, anisimovElectronEmissionMetal1974, haynesCRCHandbookChemistry2016,palikHandbookOpticalConstants1998, caldwellSubdiffractionalVolumeconfinedPolaritons2014}). Additionally, the lower sound speed in gold causes electron-boundary scattering to be the dominant contributing mechanism to the total scattering. 

Figure \ref{fig:2}d shows the trends in the plasma frequencies, where Au and Cu exhibit a characteristic downturn, in contrast to the relatively stable trend in tungsten. Gold and copper are both noble metals that conduct via their outer s-band at lower temperatures. The general trend for these noble metals is a trend upward in number density at moderate electron temperatures ($T_e < 2000$ K) due to Fermi smearing into the d-bands, resulting in a increase in $\omega_{p}$. However, at higher temperatures, the tail of Fermi smearing starts to gain access to the heavy d-bands. Hence, while the number density keeps increasing, the competing effect of the effective mass of the d-band electrons dominates the trend at $T_e > 2000$ K and thus, decreases $\omega_p$ . Tungsten, a non-noble transition metal, has a much smoother trend with electron temperature due to the d-bands participating in conduction throughout the temperature ranges simulated. 

\onecolumngrid
\newpage

\begin{figure}[t]
 \includegraphics[width=\linewidth]{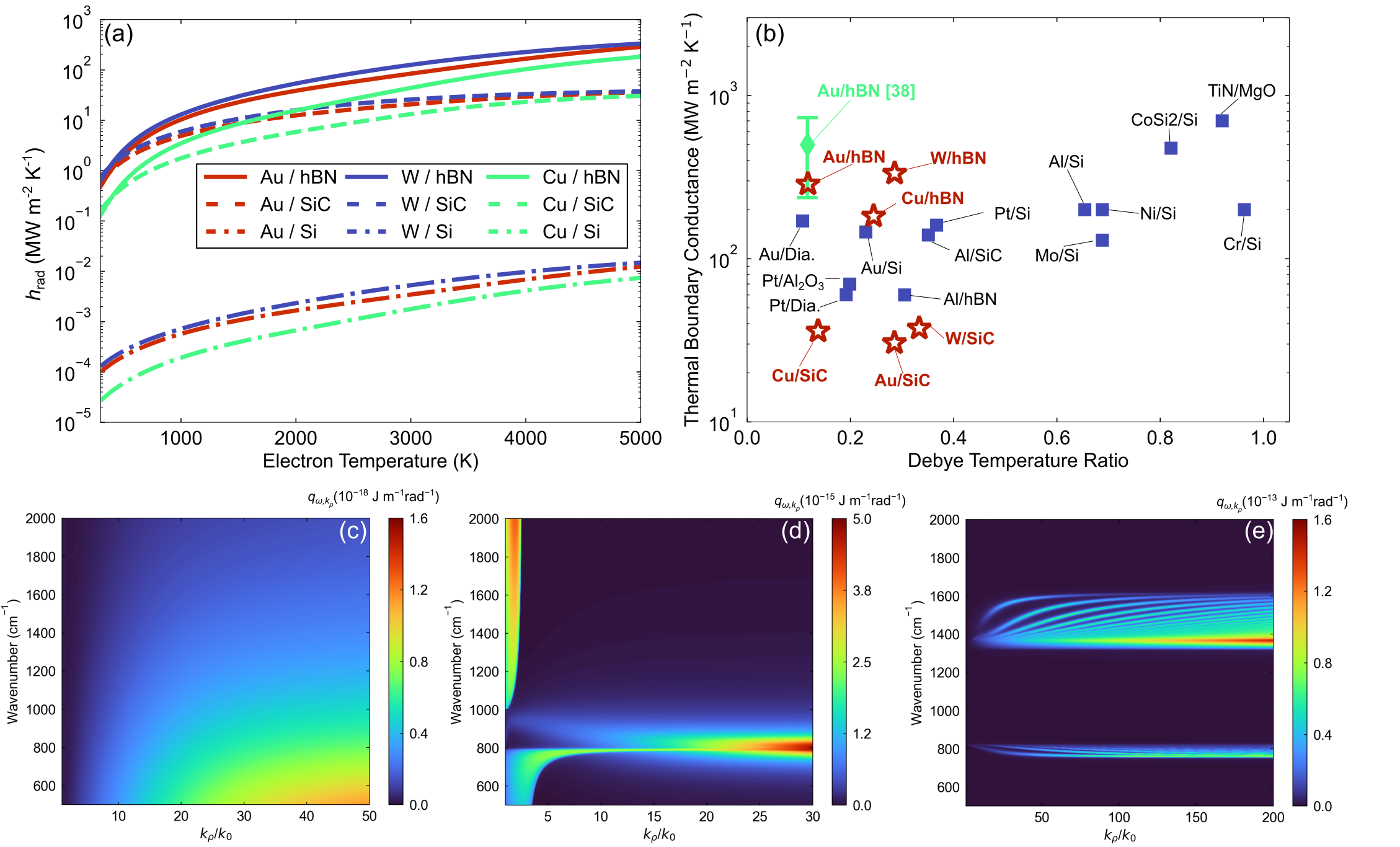}
    \newline \captionsetup{justification=raggedright,singlelinecheck=false}
    \caption[]{\textbf{} (a) FED predictions of radiative TBC at varying electron temperatures. The highest $h_{rad}$ can be expected from the absorption into high-wavevector phonon polaritons in hBN and SiC. The lack of polaritonically active optical modes in the Si results in small values of $h_{rad}$ during non-equilibrium.  (b) Comparison of non-equilibrium $h_{rad}$ calculated in this work (red stars) with measured TBC values at non-equilibrium (green diamond~\cite{hutchinsUltrafastEvanescentHeat2025}) as well as room-temperature phonon-phonon TBCs (Black squares~\cite{giriElectronPhononCoupling2020}). The values of TBC are plotted against the ratio of the film and substrate Debye temperatures, which gives a first approximation to the effective acoustic impedance matching used to estimate the efficiency of interfacial phonon transport. (c)-(e) Representative dispersion of evanescent radiative spectral heat flux per unit wavevector transferred from the metal to the substrate in the cases of 10-nm Au film with $T_e = 5000$ K on Si, 3C-SiC, and hBN, respectively. }
  \label{fig:3}
\end{figure}

\twocolumngrid

With the Drude parameters calculated at high electron temperatures, we can utilize FED to determine the radiative heat flux associated with propagating and evanescent modes across the interface between the metallic thin film and an insulator substrate~\cite{francoeurSolutionNearfieldThermal2009, salihogluEnergyTransportRadiation2020}, as described in Appendix C. This flux is then used to calculate the thermal boundary conductance with $\Delta T = T_e - T_p$. 
We considered three semi-infinite substrates (i.e., Si, 3C-SiC, and hBN) with different dielectric behaviors in the infrared region. While Si does not support any optical features that would significantly enhance the radiative heat flux, 3C-SiC and hBN can support evanescent polaritonic modes~\cite{gilesUltralowlossPolaritonsIsotopically2018} that cause a resonant increase in the heat flux. 3C-SiC is an isotropic polar wide-bandgap semiconductor with a large transverse optical (TO) absorption peak in the infrared region and supports phonon polaritons~\cite{joulainSurfaceElectromagneticWaves2005,pouriaFarfieldThermalRadiation2022,zareProbingNearFieldThermal2023}. Also, hBN is an anisotropic uniaxial medium with different in-plane and cross-plane dielectric functions. hBN possesses two hyperbolic spectral regions, in which the in-plane and cross-plane dielectric functions have opposite signs and hyperbolic phonon polaritons may be excited, resulting in a broadband enhancement of radiative heat flux~\cite{zareMeasurementNearfieldThermal2019}. The dielectric functions of these substrate materials are found using available literature parameters~\cite{palikHandbookOpticalConstants1998,gilesUltralowlossPolaritonsIsotopically2018}.

Figure \ref{fig:3}a shows the calculated results for evanescent radiative conductances, $h_{\mathrm{rad}}$, for all nine interfaces. We observe a two-order-of-magnitude increase in the interfacial radiative conductance at high electron temperatures when compared to near-equilibrium conditions (i.e., low $T_e$). The magnitude of these radiative conductances, however, is only significant for polar substrates with strong dipole oscillators supporting strong resonant modes at high wavevectors. For polaritonically active substrates, the value of $h_{rad}$ approaches the order of conductive transport  ($\sim100\mathrm{'s}$ MW m$^{-2}$ K$^{-1}$)~\cite{giriReviewExperimentalComputational2020} at electron temperatures above $3000$ K. Figure \ref{fig:3}b presents a range of literature values for total TBC measured during local thermodynamic equilibrium, i.e., $T_e=T_p=300$ K. Comparing our calculated values at $T_e = \mathrm{5000}$ K to the literature values, we see that $h_{rad}$ can be the dominant mechanism under non-equilibrium. Au and W show the highest evanescent transfer to polaritonic substrates, up to 333.7 MW m$^{-2}$ K$^{-1}$ at $T_e = 5000$ K. In the case of tungsten, this is due to the dramatically lower $\omega_p$ of the metal (see Fig. \ref{fig:2}d), resulting in a small negative dielectric function allowing more modes through the metal-insulator interface. Gold compensates for a high $\omega_p$ with overall higher scattering rates which broaden the envelope of evanescent modes and results in a comparable flux to that of W (see Section B of the Supplemental Materials for sensitivity of heat flux to each Drude term~\cite{supp}\nocite{linElectronphononCouplingElectron2008a, kresseTheoryCrystalStructures1994, anisimovElectronEmissionMetal1974, haynesCRCHandbookChemistry2016,palikHandbookOpticalConstants1998, caldwellSubdiffractionalVolumeconfinedPolaritons2014}).  

We focus on electron temperatures up to 5000 K to model scenarios of strong electron-phonon non-equilibrium, where electrons transiently reach high temperatures before thermalization with the lattice. The only measured value of radiative TBC in such non-equilibrium conditions is for Au/hBN interface, reported by Hutchins \textit{et al.}~\cite{hutchinsUltrafastEvanescentHeat2025}. The ultrafast heat transfer measurement described in Ref.~\cite{hutchinsUltrafastEvanescentHeat2025}, involving evanescent coupling between hot electrons in gold and hyperbolic phonon polaritons in hBN, aligns with the theoretical predictions presented here. The reported ``polaritonic interface conductance" of 500 MW m$^{-2}$ K$^{-1}$ at high electron temperatures agrees within uncertainty with our calculated $h_{rad}$.

To elucidate the effect of substrate on the radiative flux, we show the spectral heat flux per unit $k_\rho$ emitted by the gold film with $T_e=5000$ K into the three substrates in Figs. \ref{fig:3}c-e. The absorption from the Drude oscillator in Si results in a weak broadband heat flux. 
In the case of 3C-SiC (Fig. \ref{fig:3}d), the radiative heat flux is dominated by the excitation of bulk phonon polaritons at lower wavevectors and the TO absorption at high wavevectors. The latter occurs at a narrow spectral region resulting in a quasi-monochromatic heat flux between the metallic layer and the substrate. The transfer of flux across the interface is dictated by the allowed electromagnetic modes at each side of the interface; thus the strongest optical phonons and/or polaritons result in the most radiative flow. The metal-substrate paring that resulted in the highest $h_{rad}$ was that of W on hBN due to the intense absorption into the high-wavevector hyperbolic phonon polaritons excited within two hyperbolic regions of hBN, as illustrated in Fig. \ref{fig:3}e. These hyperbolic modes result in a directional volumetric sinking of heat from the emitting metal layer. Such enhancement of radiative interfacial heat transport under strong electron-phonon non-equilibrium has potential applications in hot electron transistors~\cite{heiblumBallisticHotelectronTransistors1990, vaziriGoingBallisticGraphene2015}, thermal switching~\cite{yangUltrafastThermalSwitching2024, thomasElectronicModulationNearField2019}, thermophotovoltaics~\cite{lenertNanophotonicSolarThermophotovoltaic2014,mittapallyNearfieldThermophotovoltaicsEfficient2021, mittapallyNearFieldThermophotovoltaicEnergy2023}, and nanophotonic devices~\cite{challenerHeatassistedMagneticRecording2009, stipeMagneticRecording152010, kimCoherentPolaritonLaser2016}. This work provides a foundation for designing advanced materials and devices that leverage these radiative mechanisms for efficient energy transduction. The presented results in Fig. \ref{fig:3} show that while the photonic energy transfer across solid-state dielectric interfaces is often overlooked when tuning the efficiency of thermal transport, the evanescent flux emitted can become significant in cases of extreme non-equilibrium.

In summary, we investigated evanescent RHT under extreme non-equilibrium in various thin film metals in contact with several dielectric absorbers. Using fluctuational electrodynamics, we predicted interfacial radiative conductance for electron temperatures ranging from 300 to 5000 K. We found that polaritonic substrate supporting surface or hyperbolic phonon polaritons in the infrared region could achieve radiative conductance comparable to its conductive counterpart. The highest conductance was observed for a tungsten film on hBN at high electron temperature due to hyperbolic phonon polaritons, surpassing 300 MW m$^{-2}$ K$^{-1}$ and rivaling  typical phonon-phonon interfacial conductances. The trends observed in radiative TBC suggest that electrons impinging on an insulating interface can emit energy not only via e-p coupling but via the transduction of thermal energy into polaritonic and photonic modes of the absorber, providing an additional pathway for thermal transport and energy transduction at the nanoscale.
\newline

\centering {\textbf{Acknowledgments}}

\justifying
This work is supported by the Office of Naval Research, Grant Number N00014-23-1-2630, and the Army Research Office, Grant Number W911NF-21-1-0119. \newline

\begin{spacing}{0.001}
\centering{\textbf{References}}
\end{spacing}

\bibliography{Biblio_Main}

\color{Black}

\onecolumngrid
 \section{End Matter}
\twocolumngrid

\justifying

\setcounter{equation}{0}
\setcounter{figure}{0}
\renewcommand{\theequation}{A\arabic{equation}}
\renewcommand{\thefigure}{A\arabic{figure}}

\noindent\textit{Appendix A: Plasma Frequency Calculation}
\newline

The plasma frequency of the metallic thin film was found as described in Eq. 2 of the manuscript, given by
\begin{equation}
     \omega_p(T_e) = \sqrt{\frac{n_e(T_e) e^2}{ 4 \pi m^*(T_e) \epsilon_0}}
\end{equation}
where $\varepsilon_0$ and $e$ are the vacuum permittivity and electron charge, respectively.  The temperature-dependent parameters in this equation are $n_e$ and $m^*$, which represent the number density and effective mass of electrons, respectively.
The number density of electrons, $n_e$, was calculated from the integration of the occupied density of states (ODOS), $g(\epsilon,T_e)$, across all energies, $\epsilon$. The ODOS was obtained by multiplying the ground state density of states (DOS), $D(\epsilon)$, by the Fermi Dirac distribution, $f_{FD}(\epsilon,T_e)$, as
\begin{equation}
 g(\epsilon,T_e) = D(\epsilon). f_{FD}(\epsilon, T_e)
\end{equation}

To find DOS for each metal film, we conducted self-consistent field calculations in the Quantum ESPRESSO package~\cite{giannozziQUANTUMESPRESSOModular2009} using a $8\times8\times8$ grid of $k$-points.  The number density of states was then calculated as
\begin{equation}
            n_e(T_e) = \int^{\infty}_{0}g(\epsilon,T_e)d\epsilon =\int^{\infty}_{0}D(\epsilon)f(\epsilon, T_e)d\epsilon
    \end{equation}

The thermal effective mass for each metal film was retrieved by the Ashcroft and Mermin formulation~\cite{ashcroftSolidStatePhysics1976} as 
\begin{equation}
m^*(T_e) = \frac{\gamma(T_e)}{\gamma_{\mathrm{free}}(T_e)}
\end{equation}

\noindent where $\gamma$ and $\gamma_{\mathrm{free}}$ are the actual and free-electron Sommerfeld Coefficients, respectively. The values of $\gamma$ and $\gamma_{\mathrm{free}}$ for each metal film at given $T_e$ is found as 
\begin{equation}
    \gamma(T_e) = \frac{C_e(T_e)}{T_e} 
\end{equation}
\begin{equation}
    \gamma_{\mathrm{free}}(T_e) = \frac{\pi^2 n_e(T_e) k_b^2}{2\mu(T_e)}
\end{equation}
    
\noindent where $C_e$ is the heat capacity of electrons, $\mu $ is the chemical potential, and $k_b$ is the Boltzmann constant. Finally, with electron effective mass and number density calculated, the plasma frequency of each metal film was found.

\mbox{}

\setcounter{equation}{0}
\setcounter{figure}{0}
\renewcommand{\theequation}{B\arabic{equation}}
\renewcommand{\thefigure}{B\arabic{figure}}

\noindent\textit{Appendix B: Electron Scattering Rate Calculation}
\newline

The electron-electron scattering rate was obtained using the Fermi liquid theory formulation, described in Ref.~\cite{gasparovElectronphononElectronelectronElectronsurface1993a}, as
\begin{equation}
\begin{split}    
    \Gamma_{ee}(T_e) = \frac{e^4 k_F^2}{16 \pi^3 \hbar^4 \varepsilon_0^2 v_F^3 q_s^2} &\left[ \frac{2k_F}{4k_F+q_s^2} + \frac{1}{q_s} \mathrm{arctan}\left(\frac{2k_F}{q_s}\right) \right]  \\ &\times \left[ \pi^2 + \left( \frac{\epsilon - \epsilon_F - \mu}{k_B T_e}\right) ^2 \right](k_B T_e)^2
\end{split}
\end{equation}

\noindent where $k_F$ and $v_F$ are the Fermi momentum and velocity respectively and are related to the Fermi energy, $\epsilon_F$, and the rest mass of the electron, $m_e$, as
\begin{equation}
k_F= \sqrt{2 m_e \epsilon_F}/\hbar
\end{equation}
\begin{equation}
v_F = k_F \hbar / m_e
\end{equation}

\noindent where $\hbar$ and $m_e$ are the reduced Planck constant and the rest mass of an electron, respectively. Also, $q_s$ represents the screening length of the electron, that is the distance at which the electrostatic force of the electron is attenuated.  The screening length in the above formulation is given by Bohm and Pines~\cite{pinesCollectiveDescriptionElectron1952} as
\begin{equation}
    q_{BP}^{-1} = \frac{4\pi \varepsilon_0 \hbar^2 1.47r_s^{1/2}}{e^2 m_e}
\end{equation}
where $r_s$ is the inter-electron distance of the simulated metal given by
\begin{equation}
    r_s = \frac{4\pi \varepsilon_0 \hbar^2 (3/4 \pi n_e)^{1/3}}{e^2 m_e}
\end{equation}

The electron-phonon scattering rate for each metal film was found by following the formula reported in Ref.~\cite{kaganovRelaxationElectronsCrystalline1957}:
\begin{equation}
    \Gamma_{ep}^{-1}(T_e) = \frac{\pi^2}{6}\frac{m^*(T_e) C_s^2 n_e(T_e)}{G_{ep}(T_e) T_e}
\end{equation}
where $C_s$ is the Debye sound speed of the metal given by
\begin{equation}
    \frac{3}{C_s^3} = \frac{1}{u_L^3} +\frac{2}{u_T^3}
\end{equation}
In Eq. (S13), $u_L$ and $u_T$ are the longitudinal and transverse sound speeds, respectively. The values of the longitudinal and transverse sound speeds for each simulated metal can be found in Table S1.

To find the electron-boundary scattering rate, we followed the formula reported in Ref.~\cite{sergeevElectronicKapitzaConductance1998} as

\begin{equation}
    \Gamma_{eb}(T_e) = \frac{3\pi}{35 \zeta(3)\Gamma_{ep}^{-1}(T_e) q_T t}\left[ 1 +2\left( \frac{u_L}{u_T}\right)^3\right]
\end{equation}

\noindent where $\zeta$ is the Riemann-Zeta function, $q_T$ is the wavevector of a thermal longitudinal phonon ($q_T = T_p / u_L$), and $t$  is the thickness of the metal film.  

\mbox{}

\setcounter{equation}{0}
\setcounter{figure}{0}
\renewcommand{\theequation}{C\arabic{equation}}
\renewcommand{\thefigure}{C\arabic{figure}}

\noindent\textit{Appendix C: Radiative Heat Flux Calculation}
\newline

Figure C1 represents a schematic of the radiative heat transfer problem under consideration as a one-dimensional layered configuration with two solid-state layers. In this configuration, layer 1 is a thin metallic film emitter (with a thickness of $t$) with a bulk vacuum layer on top, and layer 2 is a receiving dielectric half-space. The dielectric response of each layer is described using a frequency-dependent dielectric function, $\varepsilon(\omega)=\varepsilon'(\omega)+i\varepsilon''(\omega)$. As layers are infinitely long in $x$- and $y$-directions, we only consider heat flux along $z$-axis. Utilizing the framework of fluctuational electrodynamics~\cite{francoeurSolutionNearfieldThermal2009} in Cartesian coordinates, the spectral radiative heat transfer ($q_\omega$) from the thin film to the substrate is given by the time-averaged z-component of the Poynting vector, as 

\begin{figure}[t]
     \centering
     \includegraphics[width=0.75\linewidth]{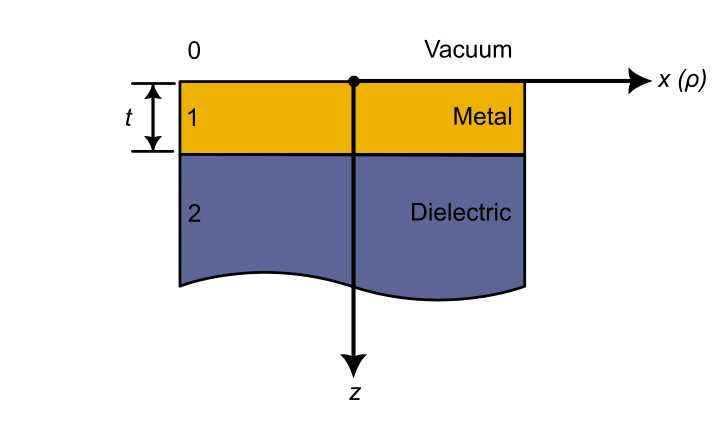}
     \captionsetup{justification=raggedright, labelfont={color=Black}}
     \caption{Schematic diagram of the configuration considered for radiative heat flux calculations.}
     \label{fig:S6}
\end{figure}

\begin{equation}
    q_{\omega} = 2 \textrm{\textbf{Re}} [<E_x H^*_y - E_y H^*_x>]
\end{equation}
\noindent where $*$ denotes the complex conjugate. Here, $E_x$ and $E_y$ ($H_x$ and $H_y$) are respectively the $x$- and $y$-components of the electric (magnetic) field $\textbf{E}$ ($\textbf{H}$), thermally emitted by the thin film.  These thermally emitted fields can be found using the fluctuation-dissipation theorem and the formalism of dyadic Green's function~\cite{francoeurSolutionNearfieldThermal2009}. By adopting a polar coordinate system and considering the azimuthal symmetry of the configuration, the radiative heat flux is then found as
\begin{equation}
\label{q_w}
\begin{split}
&q_{\omega} =  \int^{k_{\textrm{max}}}_{0} \frac{k_{0}^2\Theta(\omega,T)}{\pi^2}   \times \\ &\textrm{\textbf{Re}} \left [i  \int^{t}_{0} \sum_{\alpha = \rho,\theta,z}\varepsilon''(\omega)  (\mathbf{g}_{\rho\alpha}^E \mathbf{g}_{\theta\alpha}^{H^*} - \mathbf{g}_{\theta\alpha}^E \mathbf{g}_{\rho\alpha}^{H^*})dz\right]k_{\rho} \mathrm{d}k_{\rho}
\end{split}
\end{equation}
\noindent where $\Theta$ is the mean energy of electromagnetic states, $k_0$ is the vacuum wavevector, and $k_\rho$ is the parallel (relative to the surface) component of the wavevector. In this study, we set the upper wavevector limit $k_\max=\pi/a$, where $a$ is the lattice constant of the material. This choice reflects the physical limit imposed by the material's atomic structure and photonic response~\cite{salihogluEnergyTransportRadiation2020}.  Also, $\textbf{g}$ is the Weyl component of DGFs, given by

\begin{equation}
\label{g^E}
\textbf{g}^E(\omega,k_\rho) =  \frac{i}{2k_{z 1}}
\begin{bmatrix} 
(A^{TE}_2\widehat{\textbf{s}}\widehat{\textbf{s}}+A^{TM}_2\widehat{\textbf{p}}_2^+\widehat{\textbf{p}}_1^+)e^{-ik_{z1}z}
\\
+ (B^{TE}_2\widehat{\textbf{s}}\widehat{\textbf{s}}+B^{TM}_2\widehat{\textbf{p}}_2^-\widehat{\textbf{p}}_1^+)e^{-ik_{z1}z}
\\
+ (C^{TE}_2\widehat{\textbf{s}}\widehat{\textbf{s}}+C^{TM}_2\widehat{\textbf{p}}_2^+\widehat{\textbf{p}}_1^-)e^{ik_{z1}z}
\\
+ (D^{TE}_2\widehat{\textbf{s}}\widehat{\textbf{s}}+D^{TM}_2\widehat{\textbf{p}}_2^-\widehat{\textbf{p}}_1^-)e^{ik_{z1}z}
\end{bmatrix}
\end{equation}
\begin{equation}
\label{g^H}
\textbf{g}^H(\omega,k_\rho) =  \frac{k_2}{2k_{z 1}}
\begin{bmatrix} 
(A^{TE}_2\widehat{\textbf{p}}_2^+\widehat{\textbf{s}}-A^{TM}_2\widehat{\textbf{s}}\widehat{\textbf{p}}_1^+)e^{-ik_{z1}z}
\\
+ (B^{TE}_2\widehat{\textbf{p}}_2^-\widehat{\textbf{s}}-B^{TM}_2\widehat{\textbf{s}}\widehat{\textbf{p}}_1^+)e^{-ik_{z1}z}
\\
+ (C^{TE}_2\widehat{\textbf{p}}_2^+\widehat{\textbf{s}}-C^{TM}_2\widehat{\textbf{s}}\widehat{\textbf{p}}_1^-)e^{ik_{z1}z}
\\
+ (D^{TE}_2\widehat{\textbf{p}}_2^-\widehat{\textbf{s}}-D^{TM}_2\widehat{\textbf{s}}\widehat{\textbf{p}}_1^-)e^{ik_{z1}z}
\end{bmatrix}
\end{equation}

Here, superscript $TE$ and $TM$ refer to the transverse electric and transverse magnetic polarizations, respectively. Also, $\widehat{\textbf{s}}$ and $\widehat{\textbf{p}}_i^\pm$ are respectively the Sipe unit vectors inside $i$th layer for TE- and TM-polarizations, given by
\begin{equation}
\label{s_hat}
\widehat{\textbf{s}} =  -\widehat{\theta}
\end{equation}
\begin{equation}
\label{p_hat}
\widehat{\textbf{p}}_i^\pm =  \frac{1}{k_i}\left( \mp k_{z1}\widehat{\rho} + k_\rho \widehat{z}  \right)
\end{equation}
\noindent where $k_i=\sqrt{\varepsilon_i}k_0$ and $k_{zi}=\sqrt{k_i^2-k_\rho^2}$. Finally, the coefficients $A_2^{\gamma}$ ($B_2^{\gamma}$) and $C_2^{\gamma}$ ($D_2^{\gamma}$) represent the amplitude of waves traveling toward the positive (negative) direction of the $z$-axis in $\gamma$-polarization (where $\gamma=TE$ or $TM$) due to thermal sources emitting in the positive and negative directions of the $z$-axis, respectively. These coefficients can be found using the scattering matrix method described in Ref.~\cite{francoeurSolutionNearfieldThermal2009}.

\onecolumngrid

\end{document}


\title[SM]{Supplemental Materials for ``Interfacial Heat Transport via Evanescent Radiation by Hot Electrons''}

\setcounter{affil}{0} 

\author{William D. Hutchins}
\thanks{These two authors contributed equally.}
\author{Saman Zare}
\thanks{These two authors contributed equally.}
\affiliation{$Department~of~Mechanical~and~Aerospace~Engineering,~University~of~Virginia,~Charlottesville,~Virginia~22904, USA$}

\author{Mehran Habibzadeh}
\author{Sheila Edalatpour}
\affiliation{$Department~of~Mechanical~Engineering,~University~of~Maine,~Orono,~Maine~04469, USA$}

\author{Patrick E. Hopkins}
\email{Corresponding Author: phopkins@virginia.edu}
\affiliation{$Department~of~Mechanical~and~Aerospace~Engineering,~University~of~Virginia,~Charlottesville,~Virginia~22904, USA$}
\affiliation{$Department~of~Materials~Science~and~Engineering,~University~of~Virginia,~Charlottesville,~Virginia~22904, USA$}
\affiliation{$Department~of~Physics,~University~of~Virginia,~Charlottesville,~Virginia~22904, USA$}

\setcounter{equation}{0}
\setcounter{figure}{0}
\setcounter{table}{0}
\setcounter{page}{1}
\makeatletter
\renewcommand{\theequation}{S\arabic{equation}}
\renewcommand{\thefigure}{S\arabic{figure}}
\renewcommand{\thetable}{S\arabic{table}}
\renewcommand{\bibnumfmt}[1]{[S#1]}
\renewcommand{\citenumfont}[1]{S#1}
\titleformat*{\section}{\large\bfseries\centering}

\maketitle

\onecolumngrid

\color{Black}

\section{Section A: \textbf{Temperature Dependence of Electronic Properties}}
To find the temperature dependence of plasma frequency, the variation of $n_e$, found from Eq. A3, with electron temperature was calculated as shown in Fig. S1 across the temperature range of 300-5000 K for each metal considered in this study. We also retrieved the values of $C_e$ and $\mu$ across the simulated electron temperature range from the calculations performed by Lin et al. \cite{linElectronphononCouplingElectron2008a}, as shown in Figs. S2 and S3, respectively. Using these values, we calculated $\gamma$ and $\gamma_{free}$ at different temperatures and, consequently, the temperature dependence of electron effective mass was computed for each metal film as shown in Fig. S4. Finally, with electron number density and effective mass calculated, the plasma frequency of each metal film was found as shown in Fig. 2 of the main manuscript.

\begin{figure}[h!]
     \centering
     \includegraphics[width=0.45\linewidth]{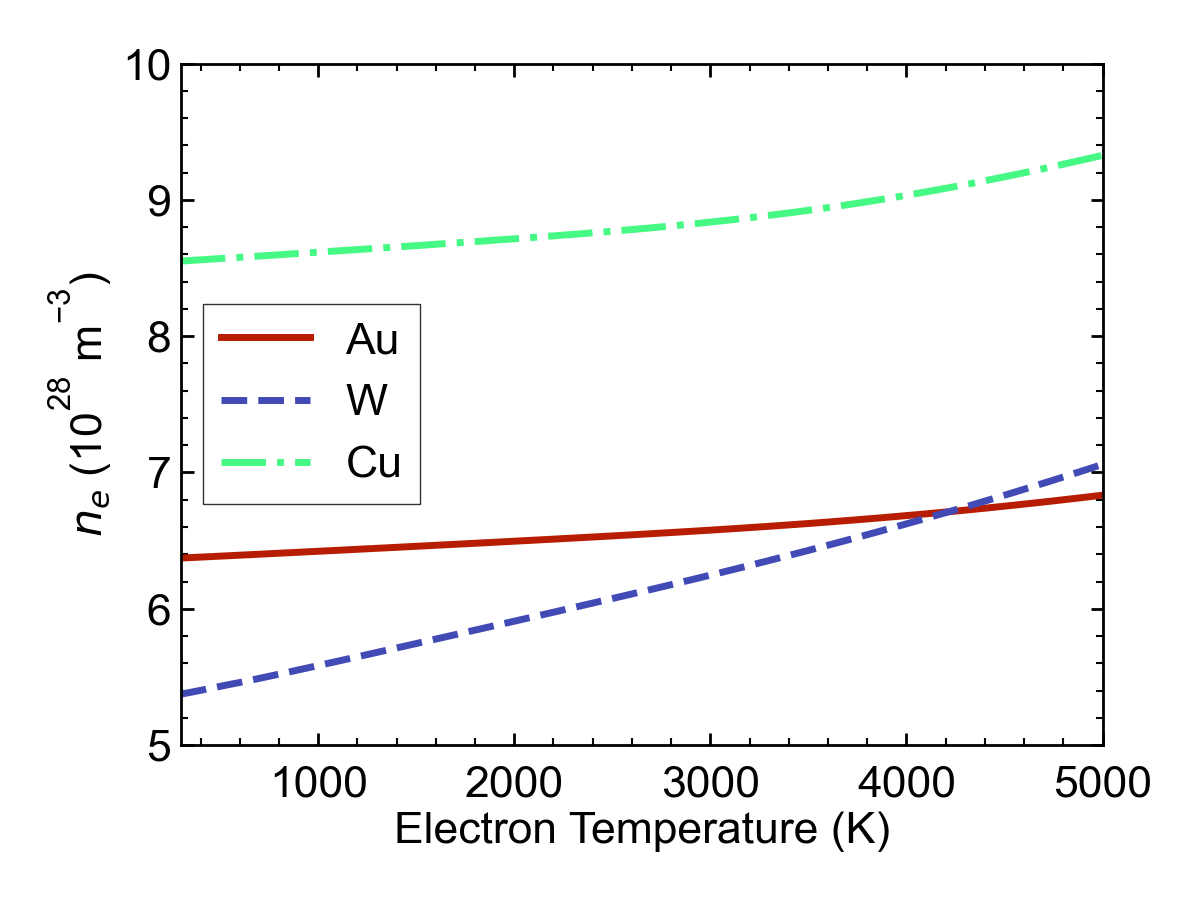}
     \captionsetup{labelfont={color=Black}}
     \caption{Variation of the electron number density with electron temperature for gold, tungsten, and copper.}
     \label{fig:S1}
 \end{figure}
 
 \begin{figure}[h!]
     \centering
     \includegraphics[width=0.45\linewidth]{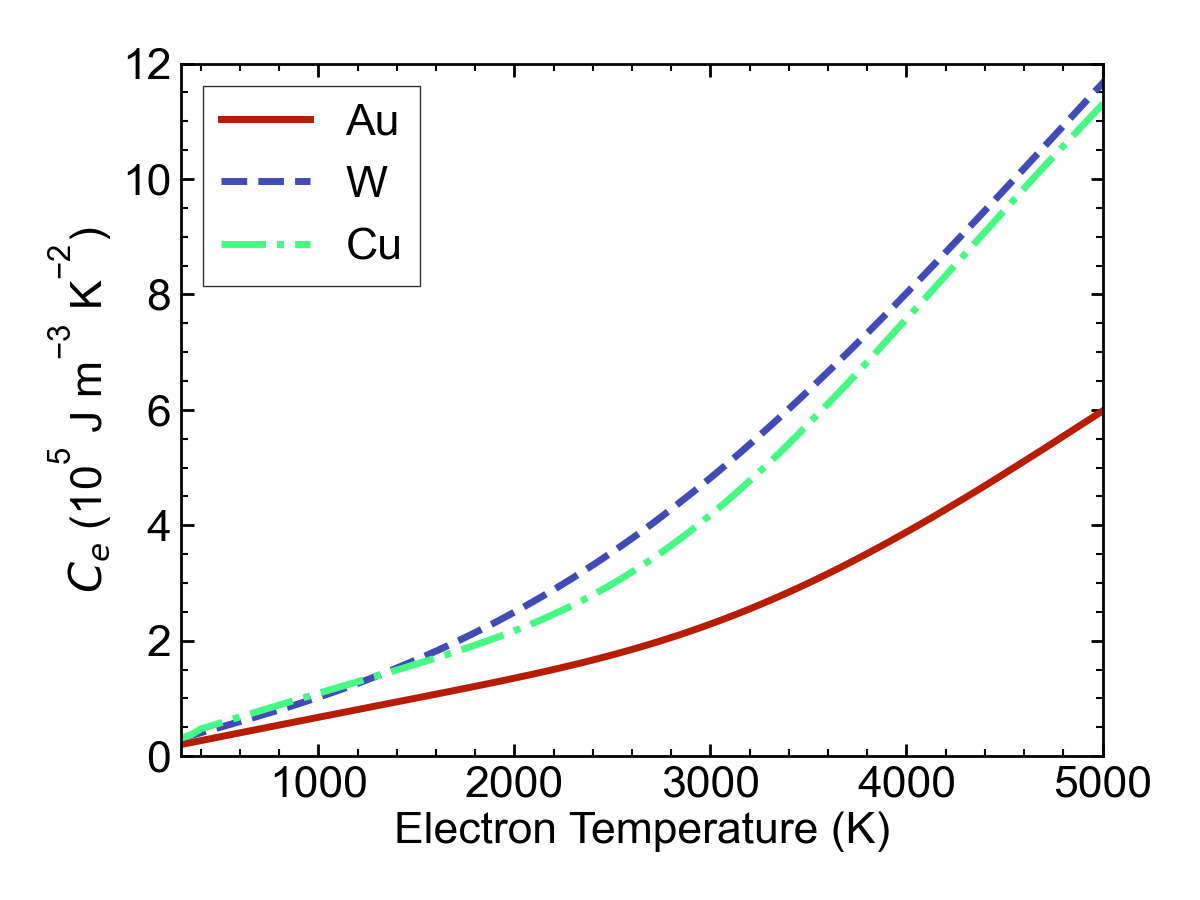}
     \captionsetup{labelfont={color=Black}}
     \caption{Temperature dependence of the heat capacity of electrons for gold, tungsten, and copper, retrieved from Ref. \cite{linElectronphononCouplingElectron2008a}.}
     \label{fig:S2}
\end{figure}

\begin{figure}[h!]
     \centering
     \includegraphics[width=0.45\linewidth]{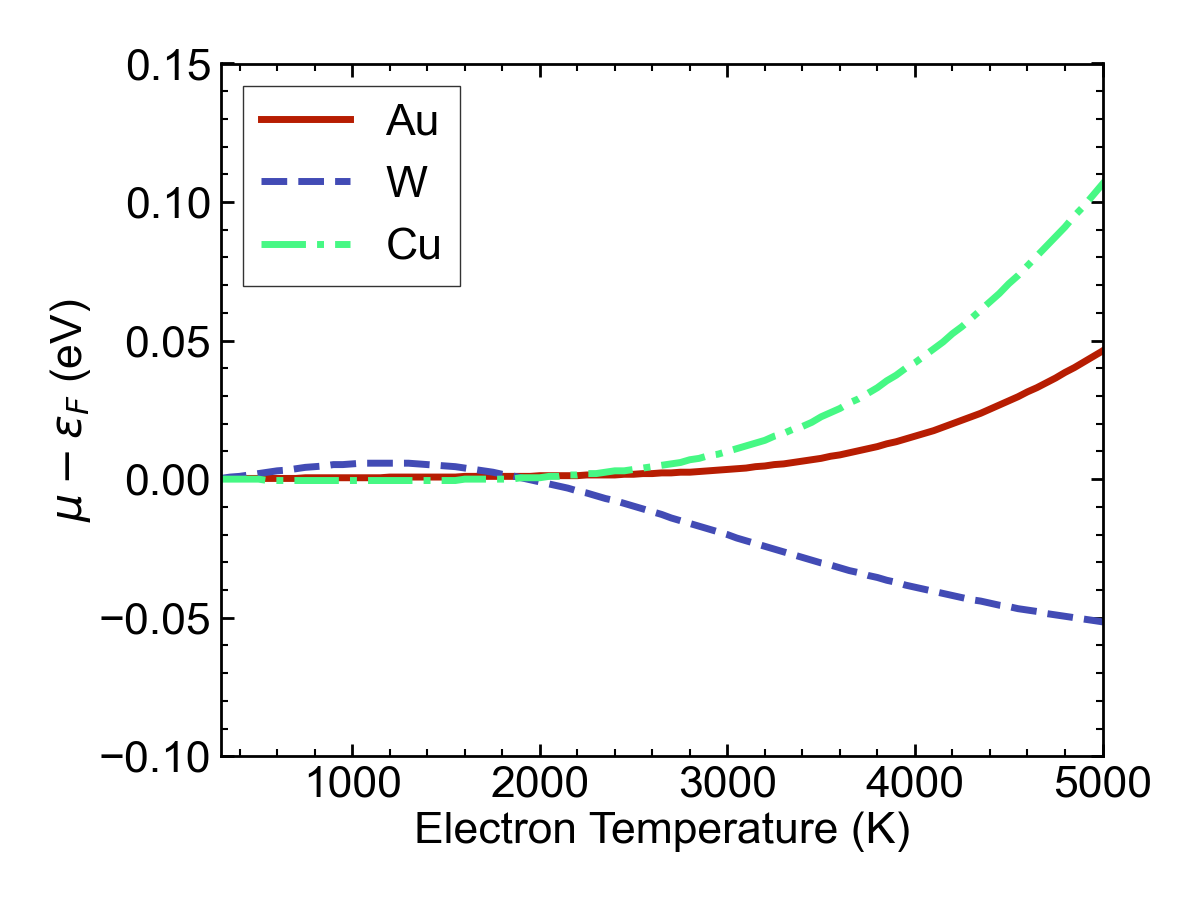}
     \captionsetup{labelfont={color=Black}}
     \caption{Temperature dependence of the chemical potential for gold, tungsten, and copper, retrieved from Ref. \cite{linElectronphononCouplingElectron2008a}. Here,  $\varepsilon_f$ is the Fermi energy.}
     \label{fig:S3}
\end{figure}

\begin{figure}[h!]
     \centering
     \includegraphics[width=0.45\linewidth]{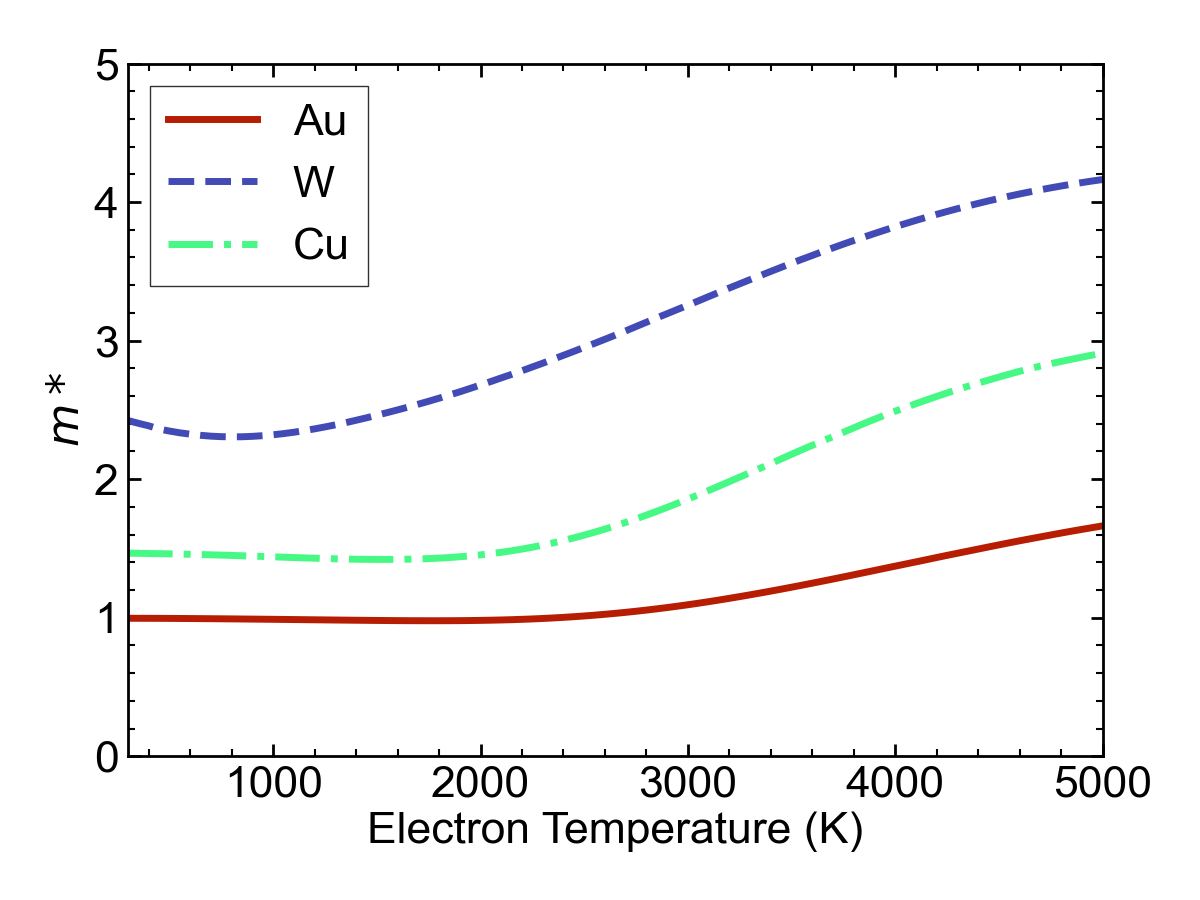}
     \captionsetup{labelfont={color=Black}}
     \caption{Temperature dependence of the electron effective mass for gold, tungsten, and copper.}
     \label{fig:S4}
\end{figure}
    
\newpage\section{Section B: \textbf{Temperature Dependence of Electron-Phonon Coupling}}
The electron-phonon coupling parameter, $G_{ep}$, was retrieved from Ref. \cite{linElectronphononCouplingElectron2008a} for each metal film and is depicted in Fig. S5. This work used  \textit{ab initio} density fluctuational theory (DFT) calculations using \textit{VASP }\cite{kresseTheoryCrystalStructures1994} at electron temperatures ranging from 0 to 5000 K. The functional relationship between $\Gamma_{ep}$ and $G_{ep}$ is reported in Ref. \cite{anisimovElectronEmissionMetal1974} to describe the case of hot electrons after laser irradiation similar to the situation depicted in Figure 1a of the main manuscript.  By inspection, we see that $G_{ep}$ is relatively constant from 0 to 2000 K only increasing significantly above this. By comparison to Fig. 2a-2c in the main manuscript, we see a slight change in curvature in this region for the calculated values of $\Gamma_{ep}$. This implies a direct relationship between $G_{ep}$ and, $\Gamma_{ep}$ as expected from Eq. (B6). However, with the similar change in curvatures found between $G_{ep}$ and $m^*$, the two effects compete and contribute to the overall reduction in curvature seen in Fig. 2 of the main text. Also, the values of the longitudinal and transverse sound speeds for each simulated metal can be found in Table S1. 

\begin{figure}[h!]
     \centering
     \includegraphics[width=0.45\linewidth]{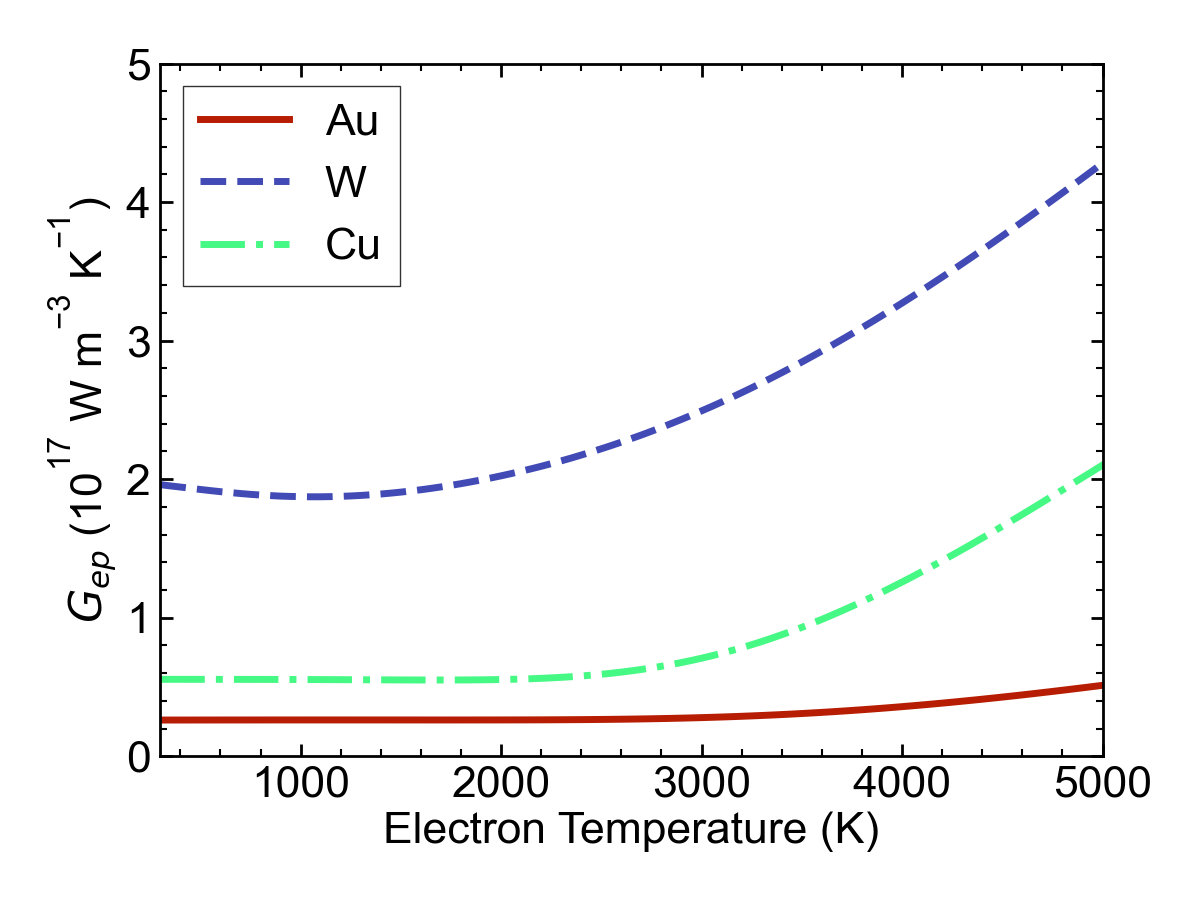}
     \captionsetup{labelfont={color=Black}}
     \caption{Temperature dependence of the electron-phonon coupling for gold, tungsten, and copper, retrieved from Ref. \cite{linElectronphononCouplingElectron2008a}. }
     \label{fig:S5}
\end{figure}

\begin{table}[h!]\centering
\color{Black}
\captionsetup{labelfont={color=Black}}
\caption{ Transverse, longitudinal, and Debye sound speeds for each metal film, retrieved from Ref. \cite{haynesCRCHandbookChemistry2016}.} 
\begin{tabular}{|p{3.5cm}|p{2cm}|p{2cm}|p{2cm}|} 
 \hline
                    & $u_T$ (m/s)&$u_L$ (m/s)&$C_s$ (m/s)\\ \hline
 Au&   1200&   3240&   1362\\ \hline
 W&   2890&   5220&   3219\\ \hline
 Cu&   2325&   4760&   2611\\ \hline
\end{tabular}
\end{table}

\newpage \section{Section C: \textbf{Optical Constants of Dielectric Substrates}}

    The isotropic dielectric function of 3C-SiC as well as the ordinary and extraordinary dielectric functions of hBN were simulated using the ``TOLO'' model as:

\begin{equation}
  \varepsilon(\omega) = \varepsilon_\infty \left( \frac{\omega_{\mathrm{LO}}-\omega^2-i\Gamma}{\omega_{\mathrm{TO}}-\omega^2-i\Gamma}\right)
\end{equation}

where $\varepsilon_\infty$, $\Gamma$, $\omega_{\mathrm{LO}}$, and $\omega_{\mathrm{TO}}$ represent the high-frequency limit to permittivity, phonon dampening, and longitudinal and transverse optical phonon frequencies, respectively. The values of these parameters for SiC and hBN were retrieved from Refs. \cite{palikHandbookOpticalConstants1998} and \cite{caldwellSubdiffractionalVolumeconfinedPolaritons2014} as listed in Table S2. For the silicon substrate, which does not support polaritonic absorption, the tabulated values of dielectric function found in \textit{Palik's Handbook of Optical Constants} \cite{palikHandbookOpticalConstants1998} were used. The dielectric functions of these substrates, used in radiative heat flux calculations, are plotted in Fig. S6.

\begin{table}[h!]\centering
\color{Black}
\captionsetup{labelfont={color=Black}}
\caption{\textcolor{Black} {Optical constants for the dielectric substrates presented in the main text retrieved from Refs. \cite{palikHandbookOpticalConstants1998} and \cite{caldwellSubdiffractionalVolumeconfinedPolaritons2014}.}}  
\begin{tabular}{|p{3.5cm}|p{2cm}|p{2cm}|p{2cm}|p{2cm}|} 
 \hline
                    & $\varepsilon_\infty$&$\omega_{\mathrm{TO}}~[\mathrm{cm}^{\mathrm{-1}}]$&$\omega_{\mathrm{LO}}~[\mathrm{cm}^{\mathrm{-1}}]$&$\Gamma ~ [\mathrm{cm}^{\mathrm{-1}}]$\\ \hline
 3C-SiC \cite{palikHandbookOpticalConstants1998}        &   6.7         &   969       &   793   &   4.8               \\ \hline
 hBN - Ordinary \cite{caldwellSubdiffractionalVolumeconfinedPolaritons2014}&   4.79         &   1617       &   1363   &   7.3               \\ \hline
 hBN - Extraordinary \cite{caldwellSubdiffractionalVolumeconfinedPolaritons2014}&   2.95         &   825       &   760   &   7.3               \\ \hline
\end{tabular}

\end{table}

\begin{figure}[h!]
     \centering
     \includegraphics[width=0.8\linewidth]{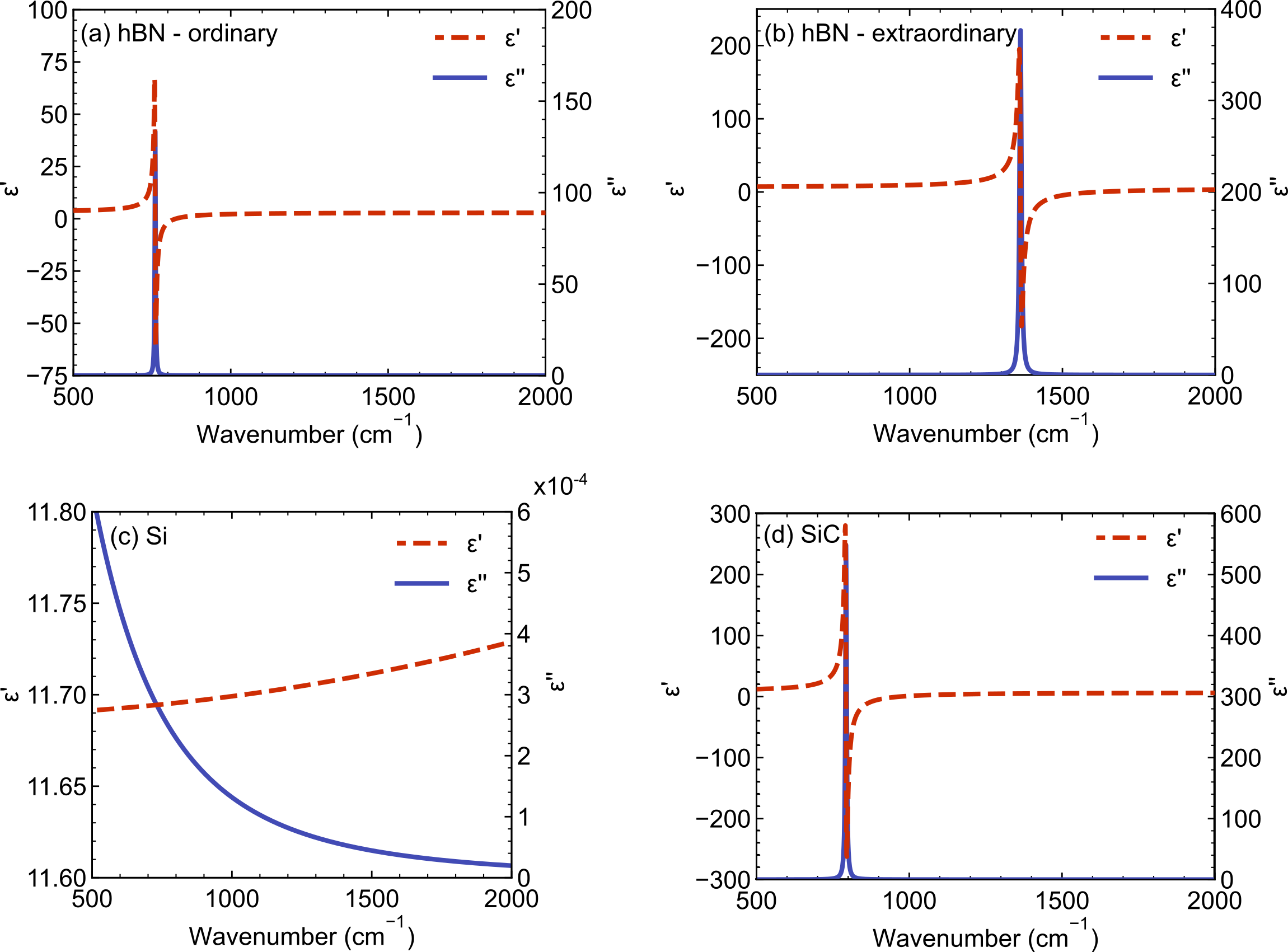}
     \captionsetup{labelfont={color=Black}}
     \caption{The dielectric functions of (a) hBN in ordinary direction, (b) hBN in extraordinary direction, (c) silicon, and (d) silicon carbide.}
     \label{fig:S6}
\end{figure}

\color{Black}

\newpage \section{Section D: \textbf{Sensitivity of $h_{rad}$ to electronic properties}}

The possible avenues for controlling radiative heat transfer at a metal/insulator interface can be explored through a sensitivity analysis. Figure S7 describes the sensitivity to each Drude parameter in the case of Au on hBN calculated by computing the change in flux for a ten percent increase in each parameter. Among all, $\omega_p$ has the strongest effect on the overall evanescent flux which causes a $\sim$15\% decrease in the flux across the interface. This strong impact is a result of the increase in available photonic modes across the interface due to the significant changes in the real part of the dielectric function with changing the plasma frequency. The scattering terms increase the total flux available in the same way, i.e., more scattering in the metal results in a broader Drude peak. The effect of increasing the thickness is that the electrons scatter less with the interface and thus, $\Gamma_{eb}$ decreases narrowing the breadth of the Drude peak.

\begin{figure}[h!]
     \centering
     \includegraphics[width=0.45\linewidth]{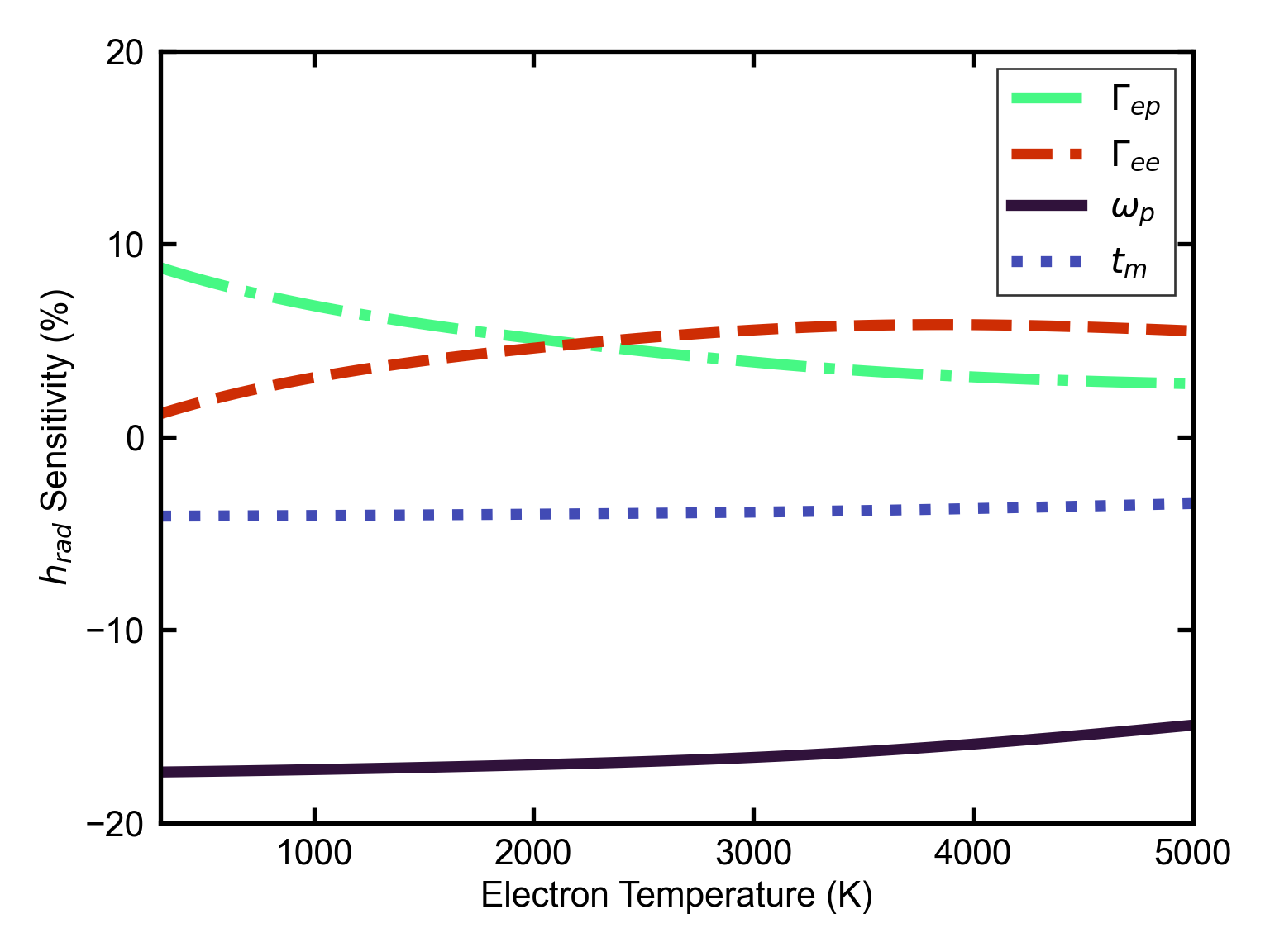}
     \caption{Sensitivity of $h_{rad}$ versus $T_e$ to each electronic property of the Drude model found for the case of Au/hBN.}
     \label{fig:S8}
\end{figure}

\begin{spacing}{0.001}
\centering{\textbf{References}}
\end{spacing}

\bibliography{Biblio_Main}